\documentclass{easychair}

\usepackage{doc}
\usepackage{makeidx}
\usepackage[leqno,fleqn,intlimits]{amsmath}
\usepackage{graphicx}
\usepackage{color}
\usepackage{float}
\usepackage{url}
\usepackage[utf8]{inputenc}
\usepackage[T1]{fontenc}

\definecolor{red_ex}{rgb}{0.7,0,0}
\definecolor{blue_ex}{rgb}{0,0,0.7}
\newcommand{\bleu}[1]{\textbf{\textcolor{blue_ex}{#1}}}

\begin{document}


\title{Beyond Stemming and Lemmatization:\\ Ultra-stemming to Improve \\ Automatic Text Summarization}


\titlerunning{Stemming is still necessary ?}

%
\author{
Juan-Manuel {\sc Torres-Moreno}\inst{1}$^,$\inst{2}
}

\institute{
  Laboratoire Informatique d'Avignon,\\
  BP 91228 84911, 
  Avignon, Cedex 09, France\\
  \email{juan-manuel.torres@univ-avignon.fr}
\and
   \'Ecole Polytechnique de Montr\'eal,\\ CP. 6128 succursale Centre-ville,
   Montréal, Québec, Canada\\
 }

\authorrunning{\sc Torres-Moreno}

\clearpage

\maketitle

\begin{abstract}
In Automatic Text Summarization, preprocessing is an important phase to reduce the space of textual representation. 
Classically, stemming and lemmatization have been widely used for normalizing words.
However, even using normalization on large texts, the curse of dimensionality can disturb the performance of summarizers. 
This paper describes a new method for normalization of
words to further reduce the space of representation. 
We propose to reduce each word to its initial letters, as a form of Ultra-stemming. 
The results show that Ultra-stemming not only preserve the content of summaries produced by this representation, but often the performances of the systems can be dramatically improved. 
Summaries on trilingual corpora were evaluated automatically with {\sc Fresa}.
Results confirm an increase in the performance, regardless of summarizer system used.
\end{abstract}

\noindent \textbf{Keywords}: Automatic Text Summarization, Lemmatization, Stemming, Ultra-Stemming

\setcounter{tocdepth}{2}

%
%

\pagestyle{empty}

\section{Introduction}
\label{sec:intro}


In Natural Language Processing (NLP), pre-processing aims to reduce the complexity of the vocabulary of the documents.
Pre-processing eliminates the punctuation, filters the function words and normalizes the morphological variants. 
In particular, the lemmatization and stemming are two commonly used techniques to normalize morphological variants.


The lexeme or word-root is the part that does not change and contains its meaning. 
The morpheme or variable part is added to the lexeme to form new words. 
Morphological analysis is a very important phase of pre-processing of NLP systems because it allows to reduce the dimension of the vector space representation in systems of  
Information Retrieval \cite{baeza,manning}.
Several applications such as Automatic Summarization, Document Indexing, Textual Classification and Question-Answering systems among others\cite{baeza}, utilize this reduction. 
However, the realization of this analysis may require the use of external resources (dictionaries, parsers, rules, etc.) which can be expensive and difficult to build, depending on language or specific domain \cite{manning}.
Some algorithms are capable to detect statistically morphological families (posed as a classification problem), 
avoiding the utilization of external resources or a priori knowledge of a language.

Automatic Text Summarization (ATS) is the process to automatically generate a compressed version of a source document \cite{torres:11}.
Query-oriented summaries focus on a user's request, and extract the information related to the specified topic given explicitly in the form of a query \cite{daume:06}.
Generic mono-document summarization tries to cover as much as possible the information content.
Multi-document summarization is a task oriented to creating a summary from a heterogeneous set of documents on a focused topic.
Over the past years, extensive experiments on query-oriented multi-document summarization have been carried out.
Extractive Summarization produces summaries choosing a subset of representative sentences from original documents.
Sentences are ordered, then assembled according to their relevance to generate the final summary \cite{mani:mayburi:99}.

This article introduces a new method of normalization of words that reduces the textual representation space, in order to improve the efficiency of 
Automatic Text Summarizers based on Vector Space Model (VSM).
We propose Ultra-stemming which reduces every word(s) to its initial(s) letter(s).
Results show that Ultra-stemming not only preserves the content of the summaries generated using this new representation, but often, 
surprisingly the performance can be dramatically improved.
To our knowledge, in summary tasks no automatic stemming method has explored this extreme possibility.
Ultra-stemming could be an interesting alternative for ATS of documents in languages $\pi$, where electronic linguistic resources are rare. 
In these languages, there are a notable absence of lemmatizers, stemmers, parsers, dictionaries, corpora and language resources in general 
(such as Nahuatl and other American Indian languages).
Our tests on trilingual corpora evaluated by the {\sc Fresa} algorithm confirm the increase of performance regardless of summarizer used and a big reduction of complexity in space and time required to generate summaries.
Related work is given in Section \ref{sec:relatedwork}.
Section \ref{sec:method} presents our Ultra-stemming strategies coupled with methods of Automatic Text Summarization.
Experiments are presented in Section \ref{sec:experiments}, followed by a discussion and the conclusions in Section \ref{sec:conclusion}.

\section{Related works}
\label{sec:relatedwork}

There are several morphological analysis methods \cite{Hammarstrom07unsupervisedlearning,Hammarstrom:2006}.
Examples of these algorithms are the Comparison of Graphs \cite{Grabar:1999}, the use of $n$-grams \cite{figuerola2000stemming,manning},
the search for analogies \cite{Lepage:1998}, the surface models based on rules \cite{Korenius:2004,porter}, 
the probabilistic models \cite{Creutz:2005}, the segmentation by optimization \cite{Creutz:2002,Goldsmith:2001}, 
the unsupervised learning of morphological families by ascending hierarchical classification \cite{Bernhard:2006}, 
the lemmatization using Levenshtein distances \cite{Dimitrios} or identifying suffixes through entropy \cite{medina}.
These methods are distinguished by the type of results obtained, by the identification of lemmas, stems or suffixes.
\textsc{Flemm}\footnote{\textsc{Flemm} is available in web site: \url{http://www.univ-nancy2.fr/pers/namer/Telecharger_Flemm.htm}} \cite{Namer:2000} 
is an analyzer for French which requires a text previously labeled by {\sc WinBrill}\footnote{{\sc WinBrill}  is available in web site: \url{http://www.atilf.fr/scripts/mep.exe?HTML=mep_winbrill.txt;OUVRIR_MENU=1}} 
or by {\sc TreeTagger}\footnote{\textsc{Treetagger}  is available in web site: \url{http://www.ims.uni-stuttgart.de/projekte/corplex/TreeTagger/DecisionTreeTagger.html}}.
\textsc{Flemm} produces, among other results, the lemma of each word of the input text.
\textsc{Treetagger} \cite{Schmid:1994} is a multilingual tool that allows to annotate texts with information of {\sl Parts-Of-Speech} (POS)
\footnote{The types of words are, for example, nouns, verbs, infinitives and particles.} and with information of lemmatization.
\textsc{TreeTagger} uses supervised machine learning and probabilistic methods \cite{cart84,quinlan}.
It can be adapted to other languages as long as the lexical resources and manually labeled corpora are available.
\textsc{FreeLing} is another example of a popular multilingual lemmatizer\footnote{\textsc{FreeLing}  is available in web site: \url{http://www.lsi.upc.edu/~nlp/freeling/}}.

Stemming transforms the variants of words into truncated forms.
Two popular stemming algorithms are the Porter stemming algorithm \cite{porter} and the Paice algorithm \cite{paice}.
The methods of stemming and lemmatization can be applied when the terms are morphologically similar.
Otherwise when the similarity is semantic, lexical search methods must be used.
To reduce semantic variation, some systems use long dictionaries.
Another systems use thesauri to associate words to entirely different morphological forms \cite{DBLP:journals/jasis/Paice96}.
Both methods are complementary since the stemming verifies similarities in the spelling level to infer lexical proximity, 
while the lexical algorithms use terminographic data  with links to synonyms.  \cite{JacTzo99a}.
\cite{Gelbukh04} presents an unsupervised genetic algorithm for stemming inflectional languages.
\cite{VilCabAlo2001a} proposes using morphological merged families  into a single term to reduce the linguistic variety of Spanish indexed texts.


Lexematization \cite{torres-moreno10} seeks morphological rearrangement of words belonging to the same family using automatic acquisition of morphological knowledge 
directly from the texts.
Although the constitution on morphological families may be interesting in itself, its main interest lies in the benefits it produces for use as normalization mechanism 
(instead or in addition to stemming or lemmatization) in specific application domains.
Probably the most common application domain is indexing terms in systems of Information Retrieval (IR).
In recent years there have been numerous articles analyzing in different languages the efficiency of stemming/lemmatization in IR.
In addition, significant progress has been made in IR in European languages other than English.
In particular, \cite{HolKamMonRij2004a} have evaluated corpora of CLEF evaluation campaigns
\footnote{Cross-Language Evaluation Forum, \url{http://www.clef-campaign.org/}} (eight European languages).
Their results show that morphological normalization techniques increase the efficiency of the IR systems and it can be used independently of the language.
Reduction algorithms using syntactic and morphosyntactic variations have shown a significant reduction of storage costs and management by storing lexemes rather than terms \cite{VilAloVil2008a}.
\cite{Air2006a} works on the impacts of compound words and standardization in IR, finding no significant performance differences between stemming and lemmatization.

However, the reality is that the linguistic resources necessary to establish morphological relationships without pre-defined rules are not available for all languages and
all domains, without mention the constant creation of neologisms \cite{cabre06}.
The proposed solution for the specific task of automatic summarization is the Ultra-stemming of letters.

Research in ATS was introduced by H.P. Luhn in 1958 \cite{luhn:58}.
In the strategy proposed by Luhn, the sentences are scored for their component word values as determined by tf*idf-like weights.
Scored sentences are then ranked and selected from the top until some summary length threshold is reached.
Finally, the summary is generated by assembling the selected sentences in original source order.
Although fairly simple, this extractive methodology is still used in current approaches.
Later on, \cite{edmundson:69} extended this work by adding simple heuristic features of sentences such as their position in the text or some key phrases indicating the importance of the sentences.
As the range of possible features for source characterization widened, choosing appropriate features, feature weights and feature combinations have became a central issue.
A natural way to tackle this problem is to consider sentence extraction as a classification task.
To this end, several machine learning approaches that uses document-summary pairs have been proposed \cite{kupiec:95,teufel:moens:97}.

\section{Pre-processing and Ultra-stemming}
\label{sec:method}

The following subsections present formally the details of the corpora studied and the proposed text pre-processing method.

\subsection{Summarization Corpora Description}
\label{sec:corpora}


To study the impact of Ultra-stemming in automatic summary tasks, we used corpora in three languages: English, Spanish and French.
The corpora are heterogeneous, and different tasks are representive of Automatic Summarization: 
generic multi-document summary and mono-document guided by a subject.

\begin{itemize}
	\item Corpus in English.
	Piloted by NIST in Document Understanding Conference\footnote{\url{http://duc.nist.gov}} (DUC) the Task 2 of DUC'04\footnote{\url{http://www-nlpir.nist.gov/projects/duc/guidelines/2004.html}}, aims to produce a short summary of a cluster of related documents.
	We studied generic multi-document-summarization in English using data from DUC'04.
	This corpus with 300K words is compound of 50 clusters, 10 documents each. 
	\item Corpus in Spanish. 
	Generic single-document summarization using a corpus from the journal {\sl Medicina Cl\'inica}\footnote{\url{http://www.elsevier.es/revistas/ctl_servlet?_f=7032&revistaid=2}}, 
	which is composed of 50 medical articles in Spanish, each one with its corresponding author abstract.
	This corpus contains 125K words.
	\item Corpus in French.
	We have studied generic single-document summarization using the Canadian French Sociological Articles corpus, generated from the journal 
	{\sl Perspectives interdisciplinaires sur le travail et la sant\'e} ({\sc Pistes})\footnote{\url{http://www.pistes.uqam.ca/}}. It contains  50 sociological articles in French, each one with its corresponding author abstract.
	This corpus contains near 400K words.
\end{itemize}

Table \ref{tab:corpus} presents the basic statistics on tokens, types and characters of the three summarization corpora studied.
\begin{table}[H]
\centering
\begin{tabular}{l|rrrr}
	\hline
	\textbf{Corpus} & \bf Language & \textbf{Tokens} & \textbf{Types} & \textbf{Letters}  \\
	\hline
		\textsc{DUC'04} & English			& 294 236  & 17 780  & 1 834 167 \\
		\textsc{Medicina Cl\'inica} & Spanish 	& 125 024  & 9 657 	& 793 937 \\
		\textsc{Pistes} & French      		& 380 992  & 18 887 	& 2 590 623 \\
	\hline
\end{tabular}
\caption{Basic Statistics for the three Summarization corpora.}
\label{tab:corpus}
\end{table}
Additionally, three large and heterogeneous corpora (generated from novels, newspaper articles and news on the Internet) were created to measure statistics of each language.
These corpora contains several million tokens in English, Spanish and French.
Table \ref{tab:corpus_generic} presents basic statistics on tokens and characters of the three generic corpora.
\begin{table}[H]
\centering
\begin{tabular}{c|cc}
	\hline
	  \bf Generic Corpus & \textbf{Tokens} & \textbf{Letters}  \\
	\hline
		\textsc{English}		& 29 346 289 	& 177 717 720 \\
		\textsc{Spanish} 		& 21 445 694  & 134 461 092 \\
		\textsc{French}      	& 17 734 663 	& 111 169 782 \\
	\hline
\end{tabular}
\caption{Basic Statistics for the three Language Generic corpora.}
\label{tab:corpus_generic}
\end{table}

\subsection{Ultra-stemming}
\label{sec:ultra-stemming}

The first step to represent documents in a suitable space is the pre-processing.
As we use extractive summarization as task, documents have to be chunked into cohesive textual segments that will be assembled to produce the summary.
Pre-processing is very important because the selection of segments is based on words or bigrams of words.
The choice was made to split documents into full sentences, in this way obtaining textual segments that are likely to be grammatically correct.
Afterwards, sentences pass through several basic normalization steps in order to reduce computational complexity.
An example of document pre-processing is given in Table \ref{tab:pretraitement}.
The process is composed by the following steps: 

\begin{enumerate}
\item \textbf{Sentence splitting}: a simple rule-based method is used for sentence splitting. 
	Documents are chunked at the dot, exclamation and question mark signs. 
\item \textbf{Sentence filtering}: words are converted to lowercase and cleared up from sloppy punctuation. 
	Words with less than 2 occurrences ($f<2$) are eliminated ({\sl Hapax legomenon} presents once in a document).
	Words that do not carry meaning such as functional or very common words are removed.
	Small stop-lists (depending of language) are used in this step.
\item \textbf{Word normalization}: remaining words are replaced by their canonical form using lemmatization, stemming, Ultra-stemming or none of them (raw text). 
\item \textbf{Text Vectorization}: Documents are vectorized in a matrix $S_{[P \times N]}$ of $P$ sentences and $N$ columns, that represent the occurrences 
	of a letter (Ultra-stemming) or a word (Lemmatization/Stemming/Raw) $j$, $j=1,2,...,N$ in the sentence $i$, $i=1,2,...,P$. 
\item \textbf{Summary generation}: each summary is generated by a summarizer based on VSM. 
\end{enumerate}
For Ultra-stemming using $n=1$ ({\sc Fix$ _1 $}), the maximum dimension $N$ may be up to 32 letters. 
This generates very compact and efficient matrices, as discussed in \ref{sec:density}.

\begin{table}[h!]
\centering
\begin{tabular}{|p{.006\textwidth}|l|}
\hline

\hspace*{-0.5em} \rotatebox{90}{\hspace*{-1.8em} Original} &

\begin{minipage}{0.93\textwidth}
\vspace*{0.2em}
\small{A federal judge Monday found President Clinton in civil contempt of court for lying in a deposition about the nature of his sexual relationship with former White House intern Monica S. Lewinsky. Clinton, in a January 1998 deposition in the Paula Jones sexual harassment case, swore that he did not have a sexual relationship with Lewinsky. Clinton later explained that he did not believe he had lied in the case because the type of sex he had with Lewinsky did not fall under the definition of sexual relations used in the case. \vspace*{-0.5em} \\ 
}
\end{minipage} \\

\hline

\hspace*{-0.5em} \rotatebox{90}{\hspace*{-2em} Splitted} &
\begin{minipage}{0.93\textwidth}
\vspace*{0.2em}
\small{\textcolor{blue}{s0/}A federal judge Monday found President Clinton in civil contempt of court for lying in a deposition about the nature of his sexual relationship with former White House intern Monica S. Lewinsky. \\
\textcolor{blue}{s1/}Clinton, in a January 1998 deposition in the Paula Jones sexual harassment case, swore that he did not have a sexual relationship with Lewinsky.\\
\textcolor{blue}{s2/}Clinton later explained that he did not believe he had lied in the case because the type of sex he had with Lewinsky did not fall under the definition of sexual relations used in the case. \vspace*{-0.5em} \\ 
}
\end{minipage} \\
\hline

\hspace*{-0.5em} \rotatebox{90}{\hspace*{-2em} Stemming} &
\begin{minipage}{0.93\textwidth}
\vspace*{0.2em}
\small{\textcolor{blue}{s0/}feder judg monday found presid clinton civil contempt court lying in deposit natur sexual relationship former white hous intern monica lewinski \\
\textcolor{blue}{s1/}clinton januari deposit paula jone sexual harass case swore sexual relationship lewinski \\
\textcolor{blue}{s2/}clinton explain believ lie case type sex lewinski fall denit sexual relat case \vspace*{-0.5em} \\ 
}
 
\end{minipage} \\
\hline

\hspace*{-0.5em} \rotatebox{90}{\hspace*{-1.2em} Fix$_1$} &

\begin{minipage}{0.93\textwidth}
\vspace*{0.2em}
\small{\textcolor{blue}{s0/}f j m f p c c c c l d n s r f w h i m l\\
\textcolor{blue}{s1/}c j d p j s h c s s r l\\
\textcolor{blue}{s2/}c l e b l c t s l f d s r u c\vspace*{-0.5em} \\ 
}
\end{minipage} \\
\hline

\hspace*{-0.5em} \rotatebox{90}{\hspace*{-2em} Matrix} &

\begin{minipage}{0.93\textwidth}
\vspace*{0.2em}
\small{\texttt{\textcolor{blue}{letter:}c d e f h i j l m n p r s u w  \\
\textcolor{blue}{s0:}\hspace{6.5mm}4 0 0 0 1 1 1 2 2 1 1 1 1 0 1 \\
\textcolor{blue}{s1:}\hspace{6.5mm}2 1 0 0 1 0 2 1 0 0 1 1 2 0 0\\
\textcolor{blue}{s2:}\hspace{6.5mm}3 1 1 0 0 0 0 3 0 0 0 1 2 1 0\vspace*{-0.5em} \\ 
}
}
\end{minipage} \\
\hline

\end{tabular}
\caption{Example of some pre-processings (Stemming, Ultra-stemming and matrix generation) applied to the document NYT19990412.0403 from DUC 2006. Document is split in sentences; 
punctuation and case are removed; words are normalized.}
\label{tab:pretraitement}
\end{table}

For comparison, four methods of normalization were applied after filtering: 
\begin{itemize}
	\item Lemmatization by simple dictionary of morphological families: 1.32M words-entries in Spanish, 208K words in English and 316K in French.  
	\item Porter's Stemming, available at Snowball site: \url{http://snowball.tartarus.org/texts/stemmersoverview.html}) for English, Spanish, French among other languages.
  	\item Raw text without normalization.
  	\item Ultra-stemming: the $n$ first letters of each word.
	For example, in the case of Ultra-stemming of $n=1$ ({\sc Fix$_1$}), inflected verbs ``sing'', ``song'', ``sings'', ``singing''... or proper names ``smith'', 
	``snowboard'', ``sex'',... are all replaced by letter ``{\bf s}''. 
\end{itemize}

\subsection{Why ultra-stemming could work?}

Although this technique could be considered a brutal destruction of the lexicon, Ultra-stemming is, in fact, an extreme stemming.
That is, this truncation represents with minimum information, what we call the {\sl stem of the stem}.
In the case of Ultra-stemming with $n = 1$, the construction of the vectors-phrases is performed in a space of $j = 1,2, ... $ 32 
classes, which produces a dense vector representation.

\begin{figure}[H]
	\centering
	\includegraphics[width=0.75\textwidth]{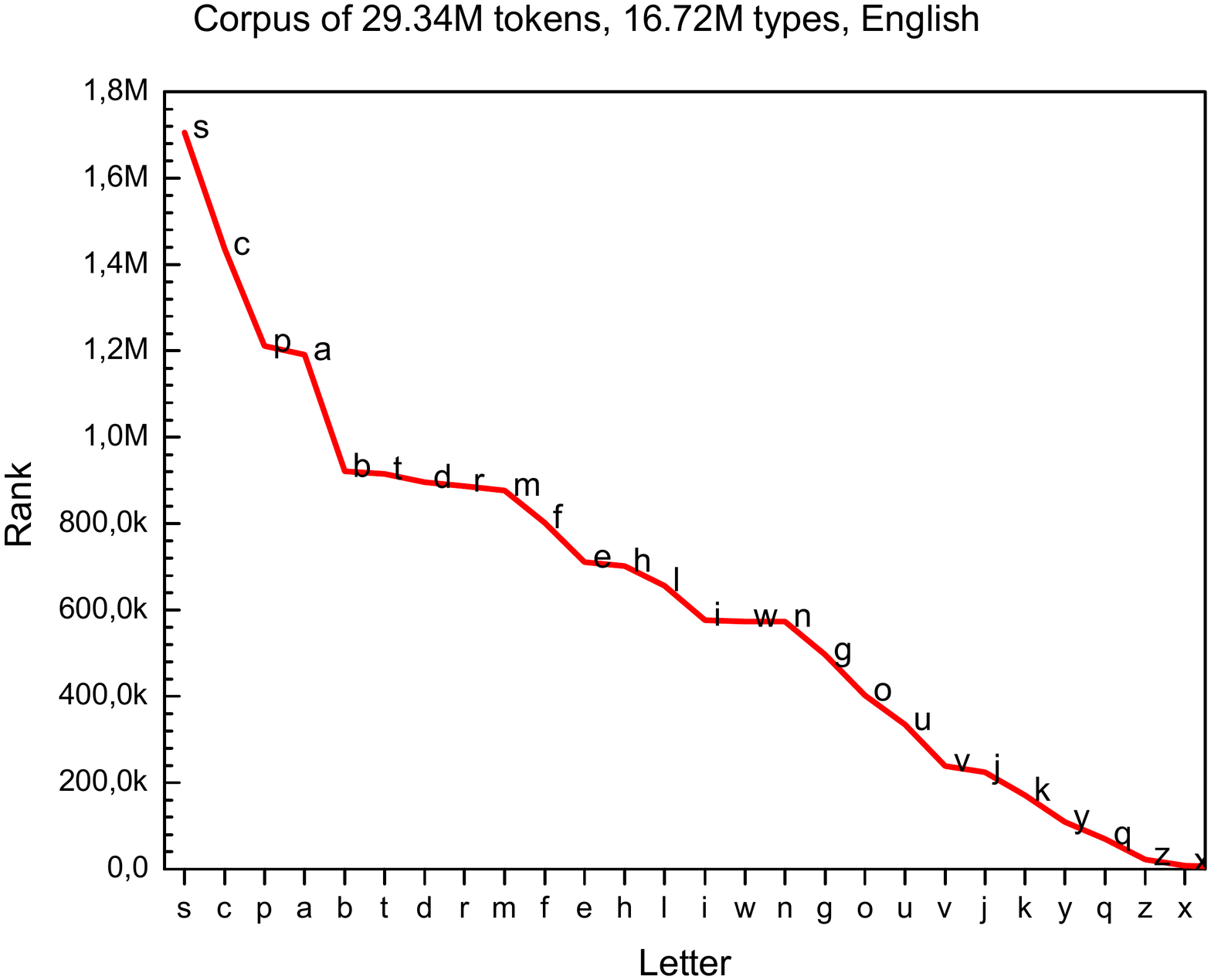}
	\caption{Scatter plot of first letter ranking for the English corpus. 
	There are 16.72M of types, after filtering of functional words and punctuation.}
	\label{fig:rankig_letters_en}
\end{figure}
\begin{figure}[H]
	\centering
	\includegraphics[width=0.75\textwidth]{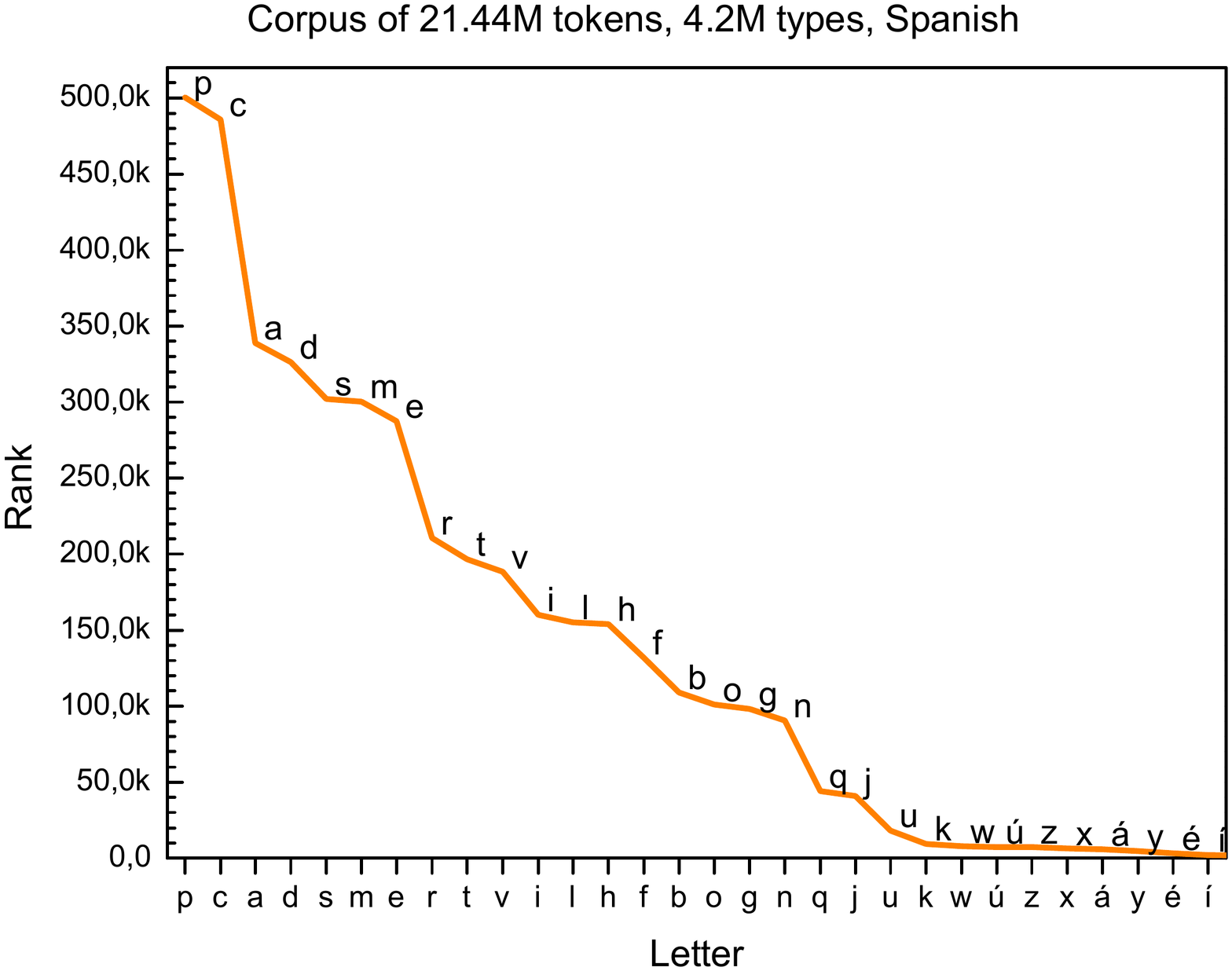}
	\caption{Scatter plot of first letter ranking for the Spanish corpus. 
	There are 4.53M of types, after filtering of functional words and punctuation.}
	\label{fig:rankig_letters_es}
\end{figure}
\begin{figure}[H]
	\centering
	\includegraphics[width=0.75\textwidth]{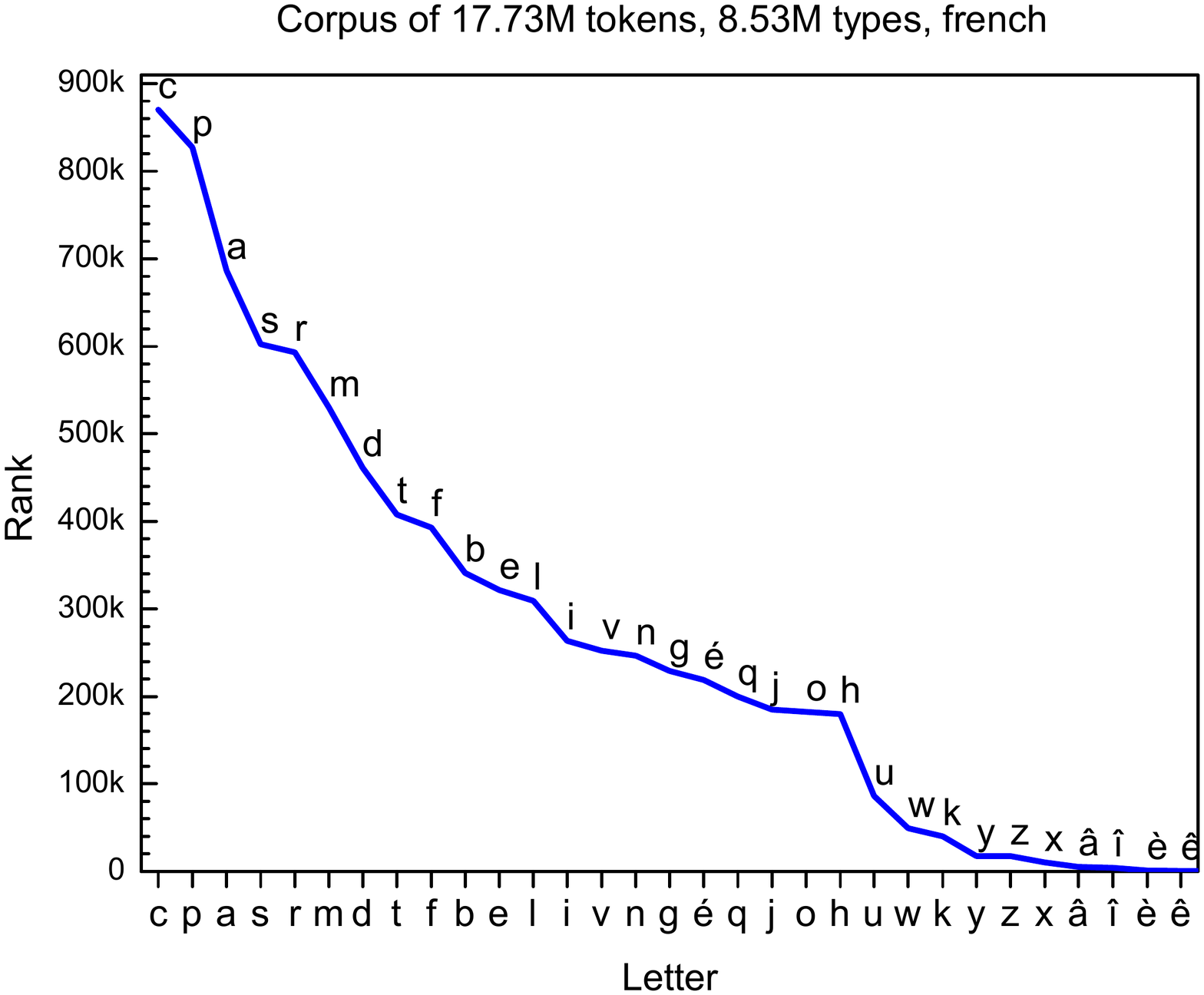}
	\caption{Scatter plot of first letter ranking for the French corpus. 
	There are 8.53M of types, after filtering of functional words and punctuation.}
	\label{fig:rankig_letters_fr}
\end{figure}

Of course, classes are not equally populated. 
Figures \ref{fig:rankig_letters_en} to \ref{fig:rankig_letters_fr} show the ranking of letters of three corpora in English, Spanish and French. 
The numbers and function words were previously removed.

In an automatic extractive summarizer, the weight of phrases is represented in a suitable vector space. 
However, if the representation is too large, the resulting representation is very sparse, which can difficult the weighting of the sentences. 
Two hypotheses are the basic ideas for using Ultra-stemming in automatic summarization task.

The first hypothesis is that a more condensed, but retaining important information of the original representation, would enable a more effective weighting for phrases extraction.
Ultra-stemming produces an extremely compact representation of documents, in a Vector Space that can reach only thirty letters, using the representation of one letter per word.
One way of evaluating the efficacity of a vector representation can be by calculating the density of the resulting matrix.
This point will be discussed in detail in the next section.
The other way is to show that two matrices $A$ and $B$ are equivalent in the sense that they contain a number of similar informations.
If $A<B$, and $A$ and $B$ represent approximately the same information, then it may be preferable to use the representation given by $A$ instead of $ B $.

Now, how does one know that two matrices contain about the same information?
The second hypothesis is that if the matrices $A$ and $B$ are correlated, then they probably represent similar information.
This point will be proved in Section \ref{sec:mantel} by the Mantel statistic test.

\subsection{Matrix density}
\label{sec:density}


Pre-processing and vectorization of documents will produce very sparse matrices.
However the density of matrices generated is directly dependent on pre-processing algorithm used.
Intuitively, the density of matrices generated by Ultra-stemming must be much greater than those generated by classical normalizations.
We have calculated the density $\delta$ of a matrix $S_{[P \times N]}$, of $P$ phrases and a vocabulary of $N$ words as a fraction of occurrences $C_w$  
of the word $w$ (elements other than 0), divided by the size of the matrix $\rho=P \times N$.
The equation \ref{eq:delta} calculates the density of $S$.
\begin{equation}
	\delta(S)=\frac{C_w > 0}{\rho}
	\label{eq:delta}
\end{equation}
This density can be an indicator of the amount of information in relation to the volume of the matrix:
lower density implies a greater amount of computation for ranking sentences.
As shown in table \ref{tab:duc_density}, the matrix produced by Ultra-stemming of letters produces a higher average density on the studied corpora.
The matrices generated by Ultra-stemming are filled approximately 50\% (56\% for English, 64\% for Spanish and 47\% for French).
The volume of the matrix generated by each pre-processing method in relation to the size of the matrix in plain text, is given by:
\begin{equation}
	V=\frac{\rho(\bullet)}{\rho(\textsc{Raw})}
	\label{eq:vol}
\end{equation} 
This volume represents a small fraction (between 5\% and 13\% depending on the language) of the matrix equivalent of plain text.


In case of the corpus {\sl Medicine Cl\'inica} the standard matrices (lemm $\approx$ 101\%, stem $\approx$ 103\%) are slightly larger than the matrix produced by the plain text (raw).
This can be explained by the presence of {\sl Hapax legomenon}.
In the case of plain text, a large number of {\sl Hapax} ($f=1 $) is eliminated and it can produce matrices slightly smaller.

\begin{table}[H]
\centering
\begin{tabular}{l|rrr|r}
	\hline
\bf DUC'04			&  				& $\langle P\rangle=238.0$  & \bf Size		  & \bf Volume V\\
\textbf{Pre-processing} 	& \textbf{Density} $\delta$ & $\langle N\rangle$ 	& $\rho=\langle P\rangle\times \langle N\rangle$ & \sc Raw=100\%\\
	\hline
\textsc{Lemmatization}	&2.6\%		& 405.5	& 96 509.0	&	96.0\%\\		
\textsc{Stemming}		&2.4\%		& 418.2	& 99 531.6	&	99.0\%\\		
\textsc{Raw}  		&2.3\%		& 424.3	& 100 983.4	&	100.0\%\\
\textsc{fix$_1$}		&\bf 55.6\%	&\bf 25.6 	&\bf 6 092.8 	&\bf	  6.0\%\\		
	\hline
	\hline
\bf Medicina Cl\'inica	&             		& $\langle P\rangle=88.6$ & \bf Size		  & \bf Volume V\\
\textbf{Pre-processing} 	& \textbf{Density} $\delta$	&  $\langle N\rangle$  &$\rho=\langle P\rangle\times \langle N\rangle$ & \sc Raw=100\%\\
	\hline
\textsc{Lemmatization}	&5.9\%  	& 177.0 	& 15 682.2	&	101.3\%\\		
\textsc{Stemming}		&5.7\%  	& 179.3 	& 15 886.0 	&	102.6\%\\		
\textsc{Raw}  		&5.1\%  	& 174.7 	& 15 478.4 	&	100.0\%\\
\textsc{fix$_1$}		&\bf 63.7\% 	&\bf 22.2 	&\bf 1 966.9 	&\bf	12.7\% \\		
	\hline
	\hline
\bf Pistes				&    		&$\langle P\rangle=319.7$ 	&  \bf Size		  & \bf Volume V\\
\textbf{Pre-processing} & \textbf{Density} $\delta$&  $\langle N\rangle$ & $\rho=\langle P\rangle\times \langle N\rangle$ & \sc Raw=100\%\\
	\hline
\textsc{Lemmatization}	& 2.0\%  	& 457.7 	& 146 326,7	&	90.0\%\\
\textsc{Stemming}		& 1.9\%  	& 474.5 	& 151 697.7	&	93.0\%\\
\textsc{Raw}  		& 1.6\%  	& 508.5 	& 162 567.5 	&     100.0\%\\
\textsc{fix$_1$}		& \bf 46.8\%  & \bf 25.0 	&\bf 7 992.5	&\bf	4.9\%\\		
	\hline
\end{tabular}
\caption{Matrix density for three corpora data. The mean dimension of matrix $S$, $\rho=\langle P\rangle\times \langle N\rangle$. 
Density $\delta(S)$ is calculated by equation \ref{eq:delta} and Volume by equation \ref{eq:vol}.}
\label{tab:duc_density}
\end{table}

Statistics for summarization DUC'04 English, {\sl Medicina Cl\'inica} Spanish  and Pistes French corpora, after removing stop-words, {\sl Hapax legomenon} and punctuation, 
are shown in table \ref{tab:stats}.
The mode of letters per word is 5, 6 and 7, and 6 respectively for each language.

\begin{table}[H]
\centering
\begin{tabular}{l|rrc|c}
	\hline
	\textbf{Corpus} &  \textbf{Words} & \textbf{Letters} & \textbf{Mean of letters} & \bf Mode on generic\\
				   &	        	&		     &	\textbf{per word}	    & \bf corpus	   \\
	\hline\hline
		\bf DUC'04  	   &			& 11 956 sentences  &			    &\bf	English	     \\
	\hline
		\textsc{Lemmatization} &   137 454  & 800 723 & 5.83	& $\bullet$	\\
		\textsc{Stemming}      &   137 101  & 764 015 & 5.57	& $\bullet$	\\
		\textsc{Raw}           &   136 582  & 902 914 & 6.61	& \bf 5	\\
		\textsc{fix$_1$}       &   137 461  & 137 461 & 1.00	& $\bullet$ \\	
	\hline\hline
	\textbf{Medicina Cl\'inica}   &           & 4 480 sentences    & 			     &\bf Spanish \\
	\hline
		\textsc{Lemmatization} &   56 063  & 484 281 & 8.64	& $\bullet$	\\
		\textsc{Stemming}      &   56 067  & 410 048 & 7.31 	& $\bullet$	\\
		\textsc{Raw}           &   56 115  & 526 660 & 9.38	& \bf 6-7	\\
		\textsc{fix$_1$}       &   56 347  &  56 347 & 1.00	& $\bullet$ \\	
	\hline	\hline
	\textbf{Pistes} 		  & 		& 16 037 Sentences   & 				&\bf French\\
	\hline
		\textsc{Lemmatization} & 167 056  & 1 505 169 & 9.01 	& $\bullet$	\\
		\textsc{Stemming}      & 167 231  & 1 264 774 & 7.56 	& $\bullet$	\\
		\textsc{Raw}           & 167 677  & 1 589 190 & 9.48	& \bf 6	\\
		\textsc{fix$_1$}       & 168 329  &   168 329 & 1.00	& $\bullet$ \\	
	\hline
\end{tabular}
\caption{Statistics for three summarization corpora  after filtering and removing punctuation.}
\label{tab:stats}
\end{table}

\begin{figure}[H]
	\centering
	\includegraphics[width=0.75\textwidth]{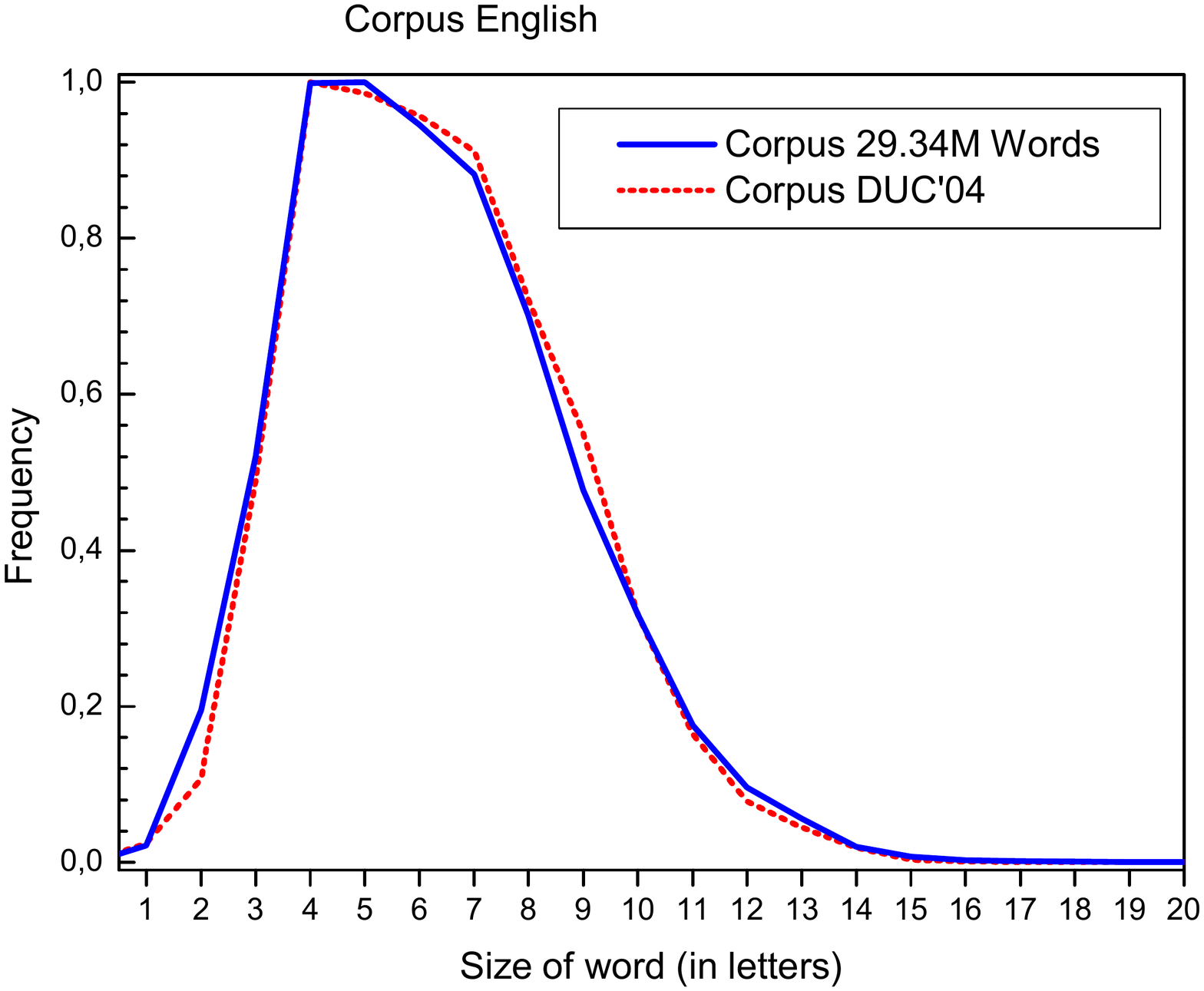}
	\caption{Scatter plot of mean length of words for two English corpora (heterogeneous and summarization raw corpora after filtering).}
	\label{fig:stats_duc}
\end{figure}
\begin{figure}[H]
	\centering
	\includegraphics[width=0.75\textwidth]{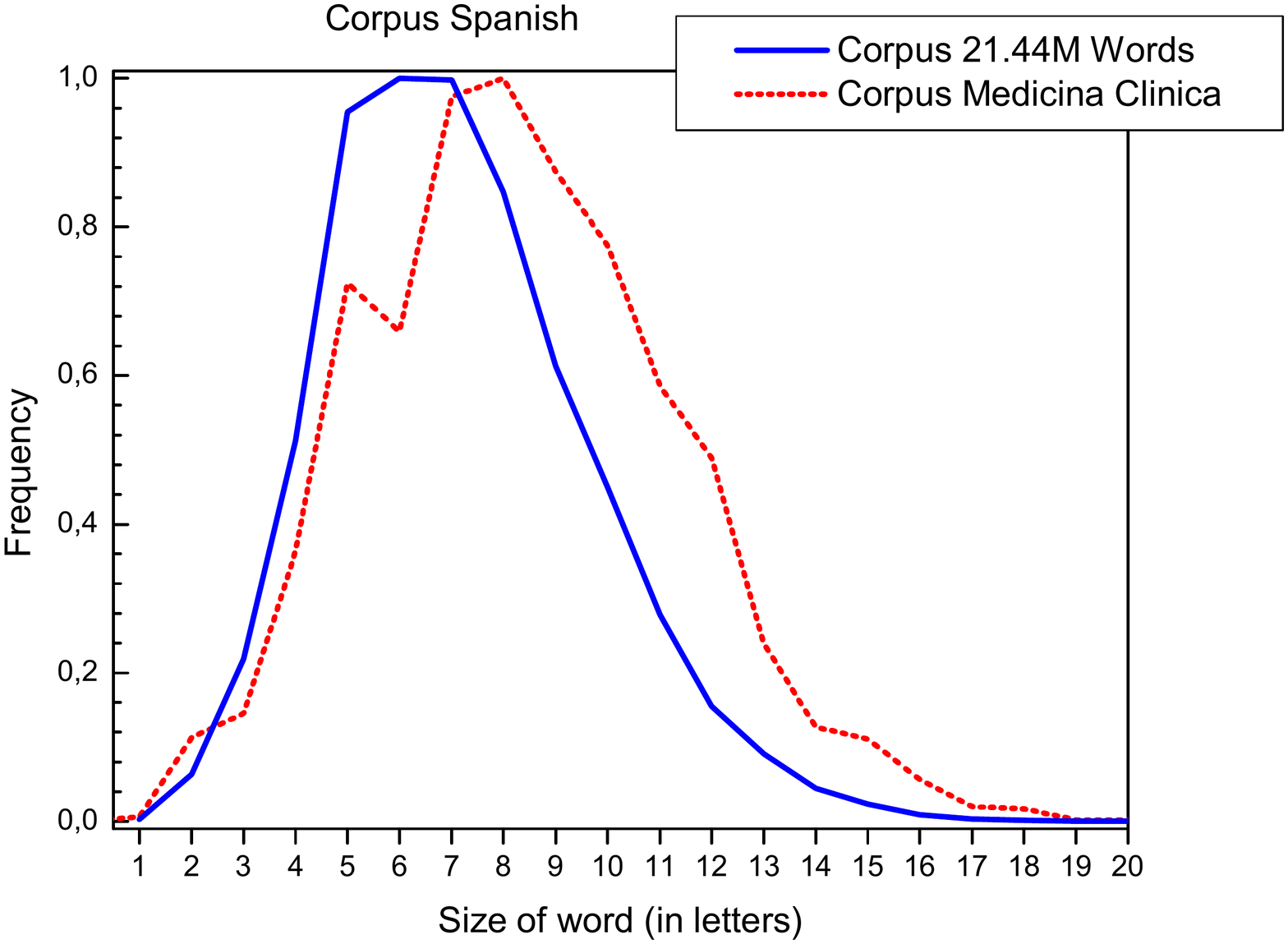}
	\caption{Scatter plot of mean length of words for two Spanish corpora (heterogeneous and summarization raw corpora after filtering).}
	\label{fig:stats_mc}
\end{figure}
\begin{figure}[H]
	\centering
	\includegraphics[width=0.75\textwidth]{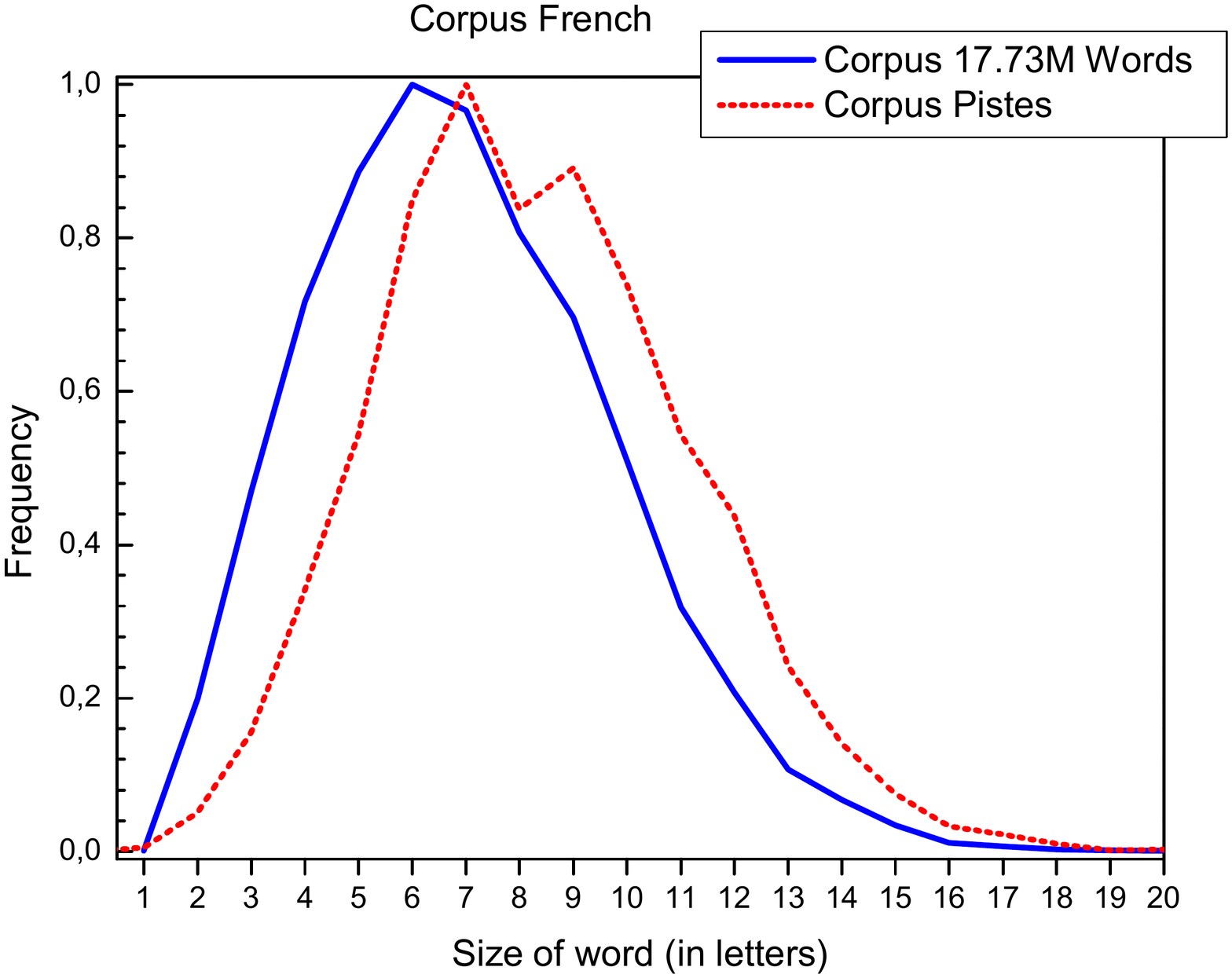}
	\caption{Scatter plot of mean length of words for two French corpora (heterogeneous and summarization raw corpora after filtering).}
	\label{fig:stats_pistes}
\end{figure}

Figures \ref{fig:stats_duc}, \ref{fig:stats_mc} and \ref{fig:stats_pistes} show the average distribution of letters per word by the three summary corpora, 
after the filtering described in \ref{sec:ultra-stemming}.
Curves are shown normalized between $[0,1]$ for the large generic and representative of the language corpora (cf Section \ref{sec:corpora}) 
and the corpora used in each of the summaries experiments.

\section{Matrix test correlation: the test of Mantel}
\label{sec:mantel}

Different methods of data analysis as ranking are based on distance matrices. 
\cite{bonnet:depeer:2002} indicates: "In some cases, researchers may wish to compare several distance matrices with one
another in order to test a hypothesis concerning a possible relationship between these matrices. 
However, this is not always evident. Usually, values in distance matrices are, in some way, correlated and therefore the usual assumption of independence between objects is
violated in the classical tests approach. Furthermore, often, spurious correlations can be
observed when comparing two distances matrices." 


As \cite{bonnet:depeer:2002} shows, in the Mantel test \cite{mantel:70}, the null hypothesis is that distances in a matrix $A$ are independent of the
distances, for the same objects, in another matrix $B$. In other words, we are testing the
hypothesis that the process that has generated the data is or is not the same in the two sets.
Then, testing of the null hypothesis is done by a randomization procedure in which the
original value of the statistic is compared with the distribution found by randomly reallocating
the order of the elements in one of the matrices.
The measure used for the correlation between $A$ and $B$ is the Pearson correlation coefficient:
\begin{equation}
	r(A,B) = \frac{1}{P-1} \sum_{i=1}^P \sum_{j=1}^P  \left[ \frac{A_{i,j} - \langle A \rangle}{\sigma_A} \right] \left[ \frac{B_{i,j} - \langle B \rangle}{\sigma_B} \right]
\label{eq:mantel}
\end{equation}
where $P$ is the number of elements in the lower (upper) triangular part of the matrix, $\langle A \rangle$ is mean for $A$ elements and $\sigma_A$ is the standard deviation of $A$ elements.

Coefficient $r>0$ measures the linear correlation and hence is subject to the same statistical
assumptions. Consequently, if non-linear relationships between matrices exist, they will be
degraded or lost ($r<0$).
The testing procedure for the simple Mantel test goes is the same of \cite{bonnet:depeer:2002}, and it is as follows:
Assume two symmetric dissimilarity matrices $A$ and $B$ of size $[P \times P]$. 
The rows and columns correspond to the same objects. 
\begin{enumerate}
	\item Compute the Pearson correlation coefficient $r(A,B)$ between the corresponding elements of the lower-triangular part of the $A$ and $B$, using equation \ref{eq:mantel}.
	\item Permute randomly rows and the corresponding columns of the matrix $A$, creating a new matrix A'.
	\item Compute $r(A',B)$ between matrices A' and B.
	\item Repeat steps 2 and 3 a great number of times. This will constitute the reference distribution under the null hypothesis.
\end{enumerate}


The calculation of the correlation between the matrix generated by the Ultra-stemming and others normalization methods is not straightforward, because the matrices are not square.
In general, the matrix produced by the Ultra-stemming have a smaller number of columns than the other ones.
Then, to calculate a correlation between matrices of different number of columns, each matrix must be converted in a symmetric matrix.


Let $S'_{[P \times N']}$ of $P$ rows and $N'$ columns be a matrix produced by Ultra-stemming,
and let $S_{[P \times N]}$ of $P$ rows and $N$  columns, be a matrix produced by a classic method of normalization such that stemming, lemmatization, etc.
We have the condition that: $ N' \le N$.
Let the new matrices be $A_{[P \times P]} = S \times S'^T$ and $B_{[P \times P]} = S \times S'^T$.
They are square symmetrical.
A standard Mantel test can indicate the degree of similarity between $A$ and $B$.
If the similarity is high ($r>0$) with a high confidence value ($p \rightarrow 0$), means that the information of the matrix $A$ is substantially the same as that 
contained in the matrix $B$.
In other words, we could replace $S'$ for $S$, for purposes of a vector representation of documents.


Tables \ref{tab:mantel_duc}, \ref{tab:mantel_mc} and \ref{tab:mantel_pistes} show the correlation of the Mantel test for the three summary corpora studied.
The correlation was calculated between the matrices $S$ generated by lemmatization (\textsc{Lemm}), stemming (\textsc{Stem}), plain text (\textsc{Raw}) 
and the matrix $S'$ generated by Ultra-stemming \textsc{Fix}$_1 $ using the initial letter.
In all cases the correlation is positive with $p$-value $<0.001$, which is significant.
The calculations were performed with the \texttt {zt} program written in $C$, of Eric Bonnet and Yves Van de Peer\footnote{zt: a software tool for simple and partial Mantel tests. 
This program can be downloaded from the Web site \url{http://bioinformatics.psb.ugent.be/software/details/ZT}} \cite{bonnet:02}.

\begin{table}[H]
\centering
\begin{tabular}{r|cccc}
	\hline
	\textbf{DUC'04} 	  & \textsc{Lemm} 	& \textsc{Stem}  & \textsc{Raw} & \textsc{fix$_1$}\\
	\hline
		\textsc{Lemm}   & $\bullet$ 	& 0.96149  	& 0.91287 	& 0.51904\\
		\textsc{Stem}   & 0.96149 		& $\bullet$  	& 0.94492 	& 0.53800\\
		\textsc{Raw}    & 0.91287 		& 0.94492  	& $\bullet$ 	& 0.46914\\
		\textsc{fix$_1$}& 0.51904 		& 0.53800  	& 0.46914 	& $\bullet$\\	
	\hline
\end{tabular}
\caption{Mantel test correlation for corpus DUC'04 data (English, $p$-value=0.001).}
\label{tab:mantel_duc}
\end{table}

\begin{table}[H]
\centering
\begin{tabular}{r|cccc}
	\hline
	\textbf{Medicina Cl\'inica} 	&\textsc{Lemm}&\textsc{Stem}&\textsc{Raw} & \textsc{fix$_1$}\\
	\hline
		\textsc{Lemm} & $\bullet$ 	& 0.96725  	& 0,91174 	& 0.58541\\
		\textsc{Stem} & 0.96725 	& $\bullet$  	& 0,91942 	& 0,49614\\
		\textsc{Raw}  & 0,91174 	& 0,91942  	& $\bullet$ 	& 0,51503\\
		\textsc{fix$_1$}& 0,58541 	& 0,49614  	& 0,51503 	& $\bullet$\\	
	\hline
\end{tabular}
\caption{Mantel test correlation for corpus {\sl Medicina Cl\'inica} data (Spanish, $p$-value=0.001).}
\label{tab:mantel_mc}
\end{table}

\begin{table}[H]
\centering
\begin{tabular}{r|cccc}
	\hline
	\textbf{Pistes} 	& \textsc{Lemm} & \textsc{Stem}  & \textsc{Raw} & \textsc{fix$_1$}\\
	\hline
		\textsc{Lemm} 	& $\bullet$ 	& 0.93016  	& 0.85708 	& 0.53801\\
		\textsc{Stem} 	& 0.93016 	& $\bullet$  	& 0,89499 	& 0.51641\\
		\textsc{Raw}  	& 0.85708 	& 0,89499  	& $\bullet$ 	& 0.45156\\
		\textsc{fix$_1$}	& 0,53801 	& 0.51641  	& 0.45156 	& $\bullet$\\	
	\hline
\end{tabular}
\caption{Mantel test correlation for corpus {\sc Pistes} data (French, $p$-value=0.001).}
\label{tab:mantel_pistes}
\end{table}


According to these correlations, in DUC'04 English corpus, the method Fix$_1$ is more correlated with Stemming normalization.
In Spanish and French corpora, Fix$_1 $ seems slightly correlated with the model lemmatization.
This is intuitively correct and according to the reduced variability of English in relation to Spanish or French.

\section{Experiments}
\label{sec:experiments}

Ultra-stemming method described in the previous section has been implemented and evaluated in several corpora in English, Spanish and French languages.
The following subsections present details of the different experiments.

\subsection{Summarizers}

Three summarization systems were used in our experiments: {\sc Cortex}, {\sc Enertex} and {\sc Artex}.
All systems have used the same text representation based on Vector Space Model, described in Section \ref{sec:ultra-stemming}.

\begin{itemize}
	\item {\sc Cortex} is a single-document summarization system using several metrics and an optimal decision algorithm \cite{torres-Moreno2001,torres:11}.
	\item {\sc Enertex} is summarization system based in Textual Energy concept \cite{fernandez:micai:07}: text is represented as a spin system where spins $\uparrow$ represents words that 
their occurrences are $f>0$ (spins $\downarrow$ if the word is not present).
	\item {\sc Artex} ({\sl AnotheR TEXt summarizer}) is a single-document summarization system that computes the score of a sentence by calculating a dot 
	product between a sentence vector and a frequencies vector, multiply by lexical used.
\end{itemize}

We have conducted our experimentation with the following languages, summarization tasks, summarizers and data sets:
	1) Generic multi-document-summarization in English with the corpus DUC'04; 
	2) Generic single-document summarization in Spanish  with the corpus {\sl Medicina Cl\'inica}  and
	3) Generic single document summarization in French  with the corpus {\sc Pistes}.

Then, we have applied the summarization algorithms following the pre-processing algorithm and 
finally, results have been evaluated using {\sc Fresa}.

\subsection{Summaries Evaluation}

To evaluate the quality of a summary is not an easy task, and remains an open question.
DUC conferences have introduced the ROUGE evaluation \cite{lin:2004rpa}, wich measures the overlap of $n$-grams between a candidate summary
and reference summaries written by humans.
In other hand, several metrics without references have been defined and experimented at DUC and TAC\footnote{\url{www.nist.gov/tac}} workshops. 
\textsc{Fresa} measure \cite{torres:10poli} is similar to \textsc{Rouge} evaluation but it does not uses reference summaries.
It calculates the divergence of probabilities between the candidate summary and the document source.
Among these metrics, Kullback-Leibler (KL) and Jensen-Shannon (JS) divergences have been used \cite{louis:nenkova:08,torres:10poli} to evaluate the informativeness of summaries.
In this paper, we use {\sc Fresa}, based in KL divergence with Dirichlet smoothing, like in the 2010 and 2011 INEX edition \cite{SanJuan:11}, to evaluate the informative content of summaries 
by comparing their $n$-gram distributions with those from source documents.

{\sc Fresa} only considered absolute log-diff between frequencies. 
Let $T$ be the set of terms in the source. 
For every $t \in T$ , we denote by $C_t^T$ its occurrences in the source and by $C_t^S$ its occurrences in the summary. 
The {\sc Fresa} package computed the divergence between the source and the summaries as:
\begin{equation}
	{\mathcal D}(T||S) = \sum_{t \in T} \left| \log \left(\frac{C_t^T}{|T|} + 1\right) - \log\left( \frac{C_t^S}{|S|} + 1\right)\right| 
\end{equation}

%
To evaluate the quality of generated summaries, several automatic measures were computed:
\begin{itemize}
	\item \textsc{Fresa$_1$}: Unigrams of single stems after removing stop-words.
	\item \textsc{Fresa$_2$}: Bigrams of pairs of consecutive stems (in the same sentence).
	\item \textsc{Fresa$_{SU4}$}: Bigrams with 2-gaps also made of pairs of consecutive stems but allowing the insertion between them of a maximum of two stems.
	\item $\langle\textsc{Fresa}\rangle=\frac{\textsc{Fresa}_1+\textsc{Fresa}_2+\textsc{Fresa}_{SU4}}{3}$ is the mean of {\sc Fresa} values, and represents the final score in our experiments.
\end{itemize}

The scores of {\sc Fresa} are normalized between 0 and 1.
High values mean less divergence regarding the source document summary, reflecting a greater amount of information content.
All summaries produced by systems were evaluated automatically using {\sc Fresa} package.

\subsection{Results}

Below we present separate results for the three languages.
In this way, we have analyzed linguistic phenomena specific to each language.

\subsubsection{English corpus}

\begin{figure}[H]
	\centering
	\includegraphics[width=0.75\textwidth]{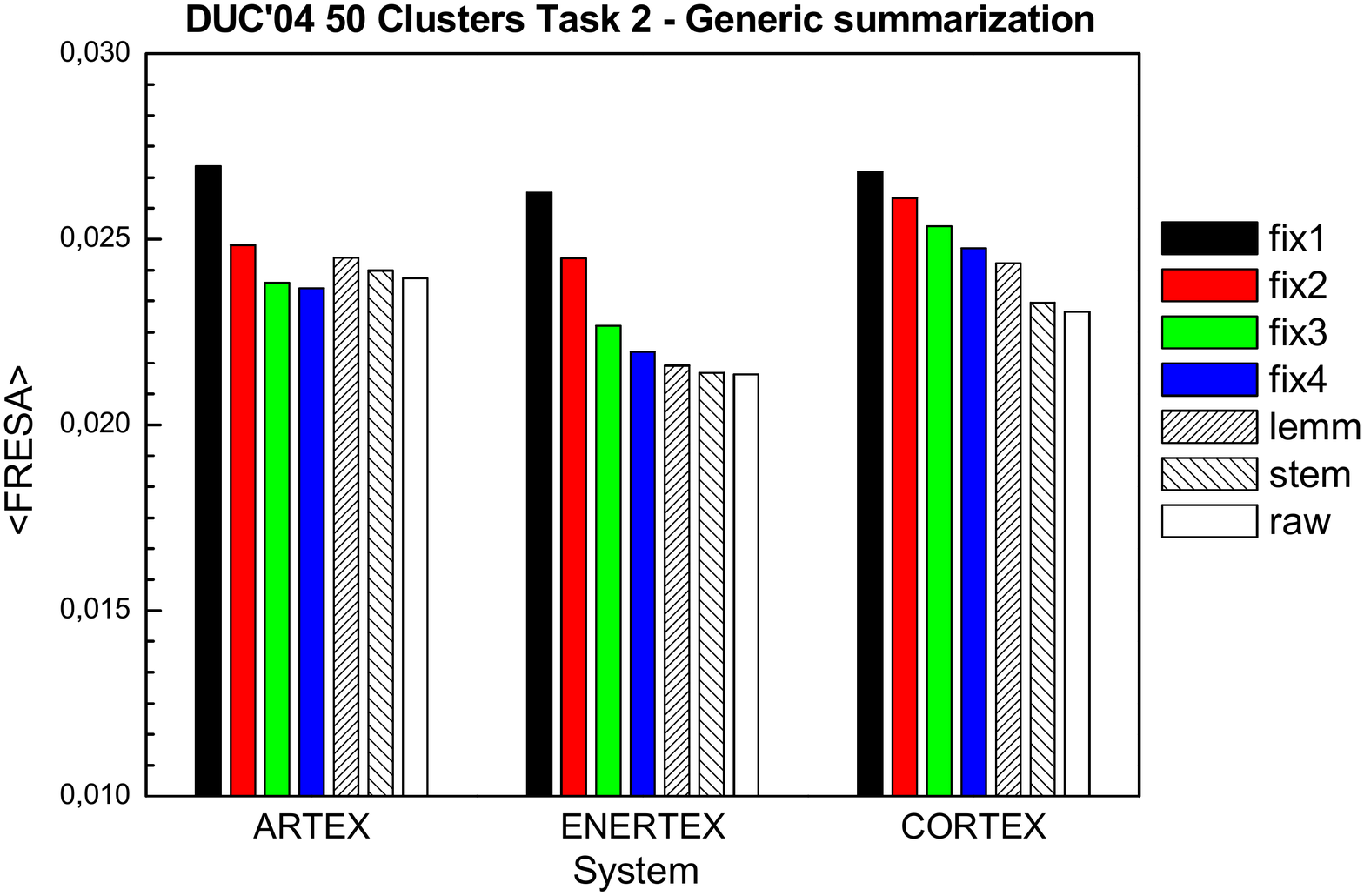}
	\caption{Histogram plot of content evaluation for corpus DUC 2004 Task 2, with $\langle$\textsc{Fresa}$\rangle$ measures, for each summarizer and each normalization.}
	\label{fig:duc04}
\end{figure}

Results in figure \ref{fig:duc04} show that Ultra-steming improves the score of the three automatic summarizer systems.
This result is remarkable for {\sc Fix$_1 $}, whose average matrix represents only 6\% of the matrix volume in plain text.

\begin{figure}[H]
	\centering
	\includegraphics[width=0.75\textwidth]{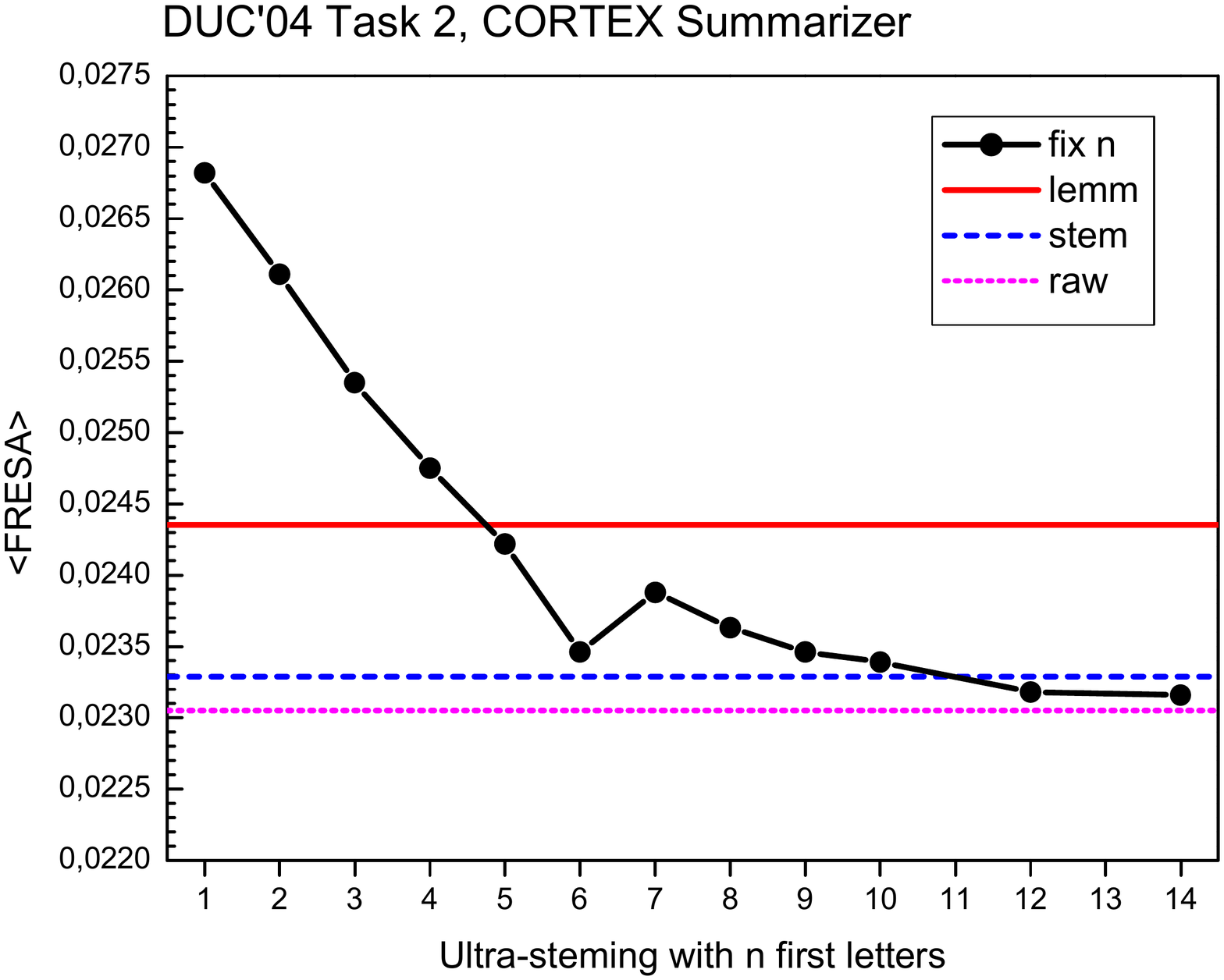}
	\caption{Scatter plot of $\langle${\sc Fresa}$\rangle$ mean of Ultra-stemming using $n$ first letters (corpus DUC 2004 Task 2, {\sc Cortex} summarizer).}
	\label{fig:duc04_fixn}
\end{figure}

As shown in Figure \ref{fig:duc04}, the performance of the three summarizers is improved using the Ultra-stemming in relation to other normalizations.
So, in particular, using lemmatization (the best score between the two classic normalizations), the summarizer {\sc} Artex, goes from 0.02451 to 0.02697 
using normalization {\sc Fix$_1 $}, i.e. an increase of 10\%.
{\sc Cortex} increases of 0.02435 to 0.02682, an augmentation of 10.1\% and summarizer {\sc Enertex} increases of 0.02141 to 0.02626, an augmentation of 22.7\%.


A detailed analysis for a particular summarizer is shown in Figure \ref{fig:duc04_fixn}.
This figure shows the average score {\sc Fresa} obtained on DUC'04 English corpus, in function of Ultra-stemming used, of $n=1,2,..., 14$ letters, 
for the automatic summarizer {\sc Cortex}.
By comparison, the values {\sc Fresa} for lemmatization (lemm), stemming (stem) and plain text (raw) are shown in the graph.

\subsubsection{Spanish corpus}

\begin{figure}[H]
	\centering
	\includegraphics[width=0.75\textwidth]{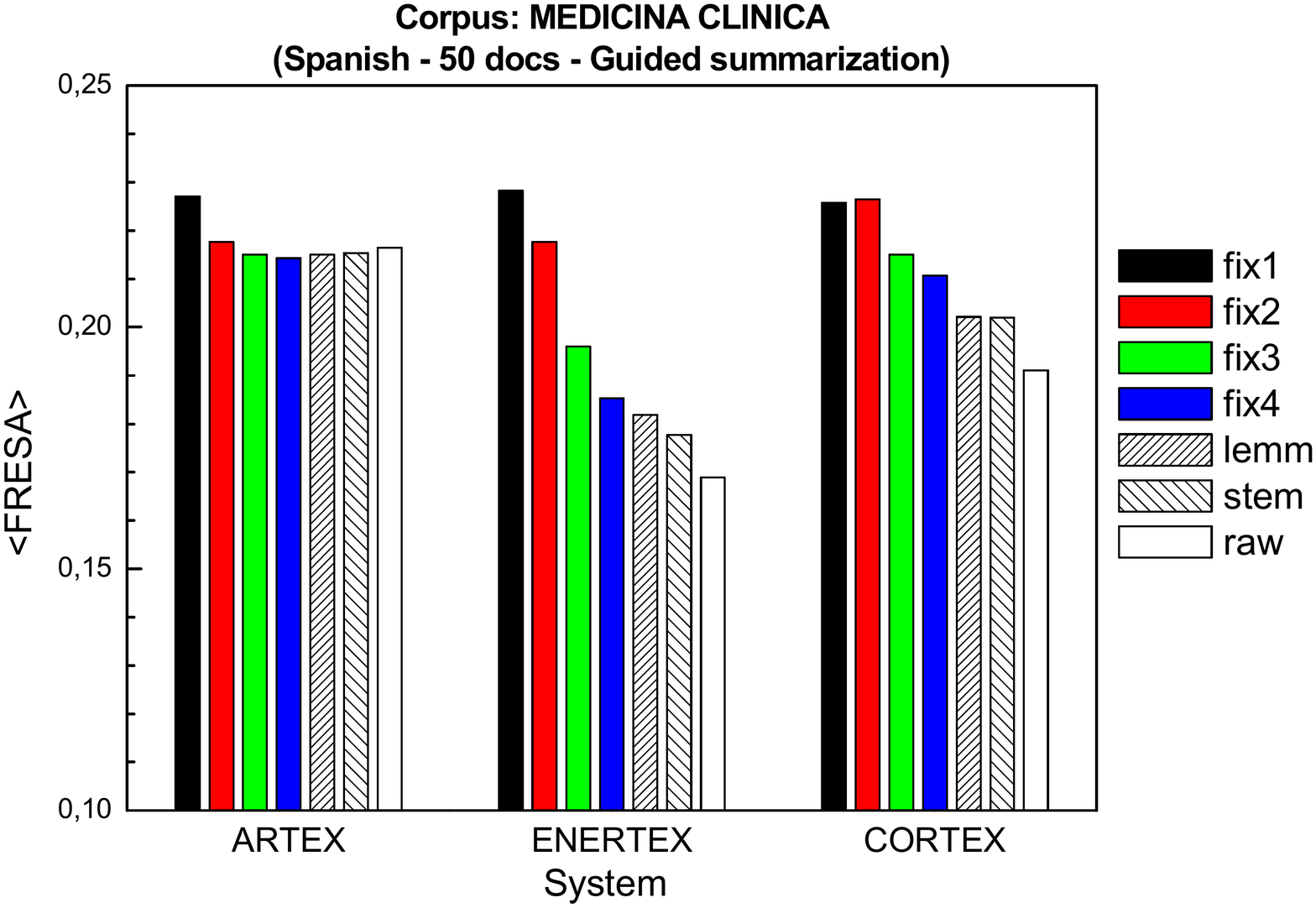}
	\caption{Histogram plot of content evaluation for Spanish corpus {\sl Medicina Cl\'inica} with $\langle$\textsc{Fresa}$\rangle$ scores for each summarizer.}
	\label{fig:medicina}
\end{figure}

Spanish is a language with greater variability than English.
Results in figure \ref{fig:medicina} shown that ultra-steming  improves the score of the three systems of automatic summarization utilized.
In the case of summarizers {\sc Cortex} and {\sc Artex}, stemming and lemmatization substantially obtains the same scores, which does not occur with {\sc Enertex}.
However, comparing Ultra-stemming against stemming {\sc Fix$_1$}, the three summarizers are benefiting of an increased score ({\sc Artex} 5\%, {\sc Enertex} 5.25\% 
and {\sc Cortex} 7.11 \%).


\begin{figure}[H]
	\centering
	\includegraphics[width=0.75\textwidth]{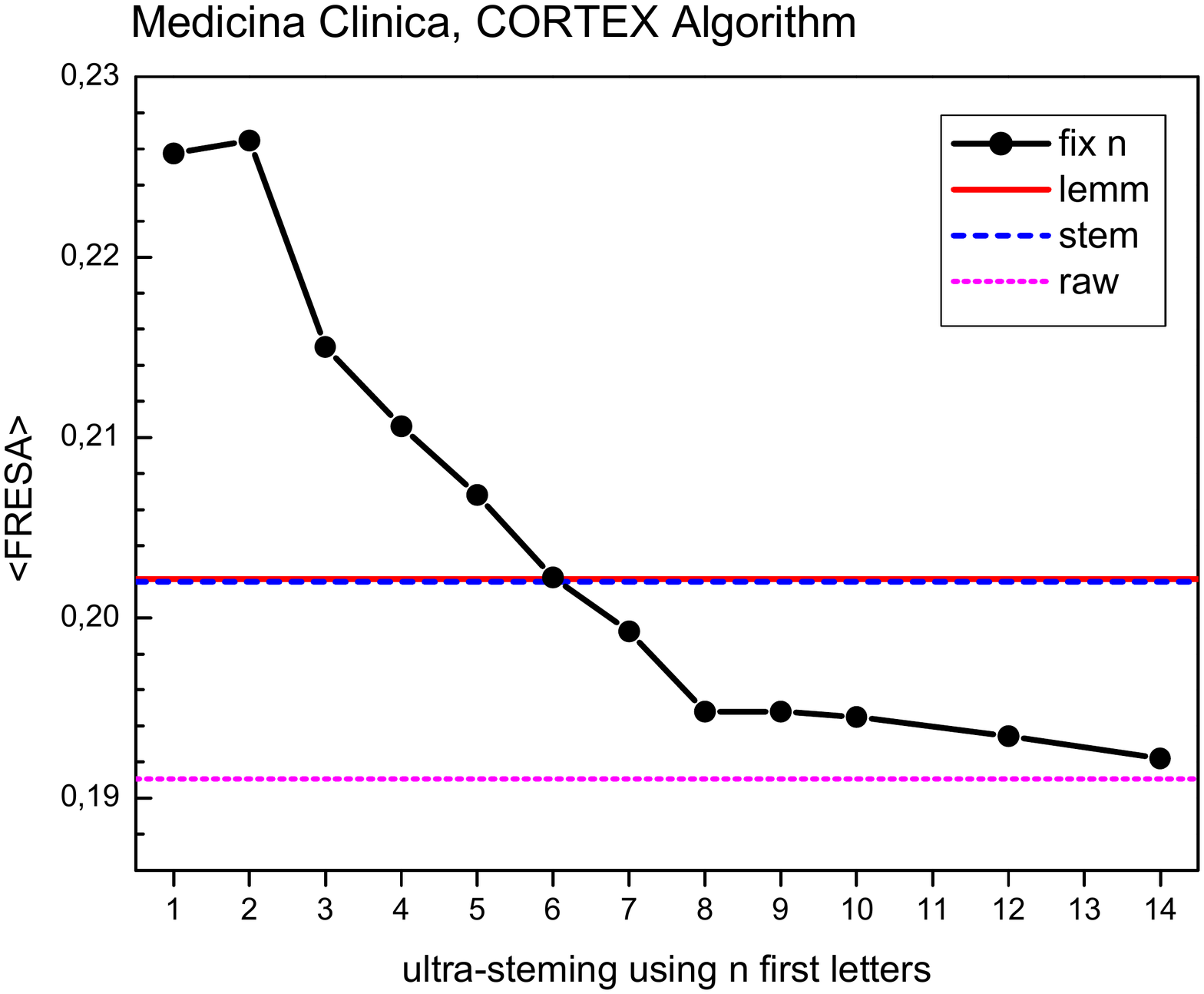}
	\caption{Scatter plot of $\langle${\sc Fresa}$\rangle$ mean vs. Ultra-stemming using $n$ first letters (corpus {\sl Medicina Cl\'inica}, {\sc Cortex} summarizer).}
	\label{fig:mc_fixn}
\end{figure}

Figure \ref{fig:mc_fixn} shows the mean score $\langle$\textsc{Fresa}$\rangle$ on the Spanish corpus {\sl Medicine Cl\'inica}, based on the ultra-stemming ($n=1,2,...14$ 
letters) using automatic summarizer {\sc Cortex}.
Values {\sc Fresa} for lemmatization ({\sc Lemm}), stemming ({\sc Stem}) and plain text ({\sc Raw}) are also shown.

\subsubsection{French corpus}

\begin{figure}[H]
	\centering
	\includegraphics[width=0.75\textwidth]{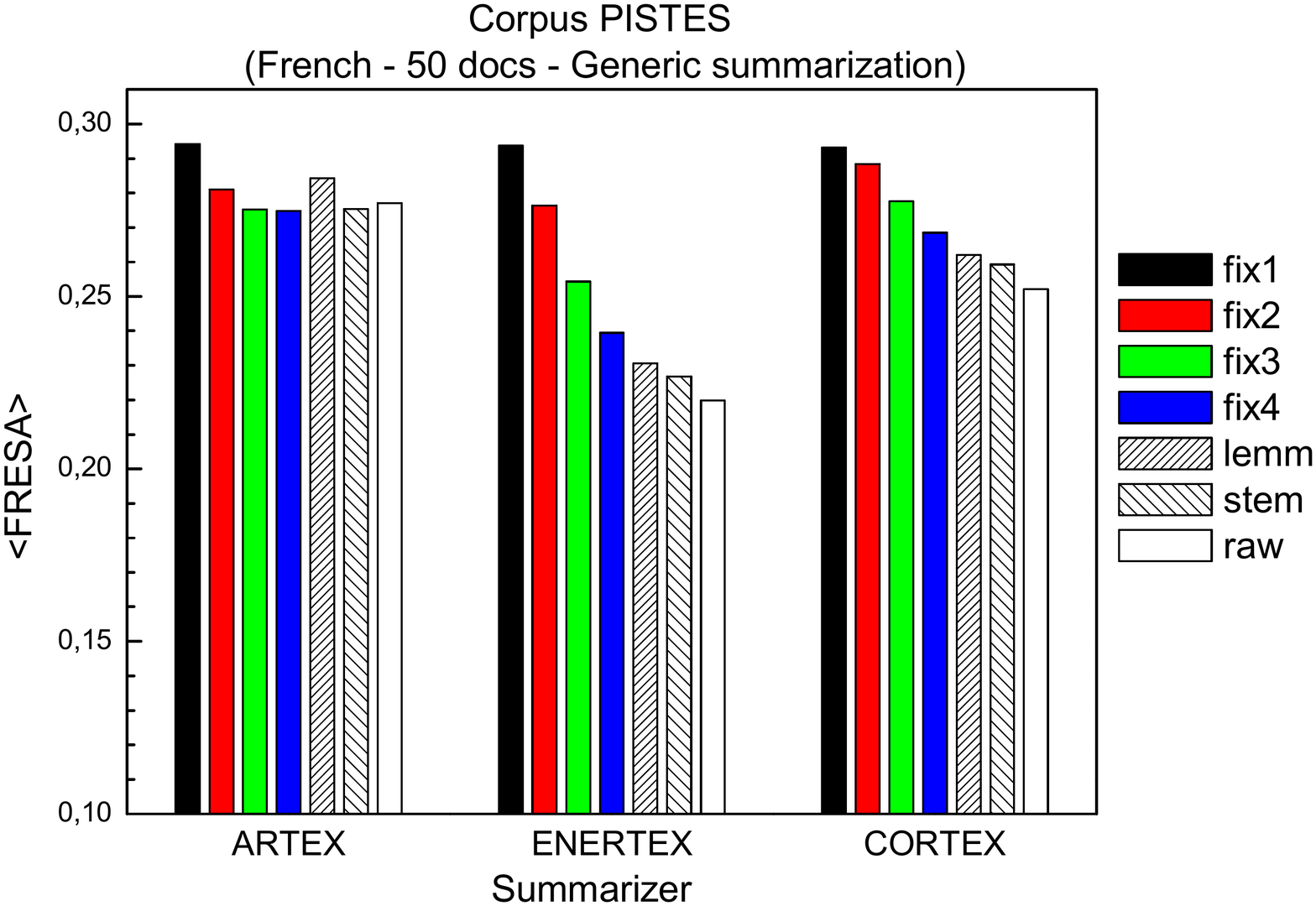}
	\caption{Histogram plot of content evaluation for French corpus {\sc Pistes} with {\sc Fresa} scores for each summarizer.}
	\label{fig:pistes}
\end{figure}

\begin{figure}[H]
	\centering
	\includegraphics[width=0.75\textwidth]{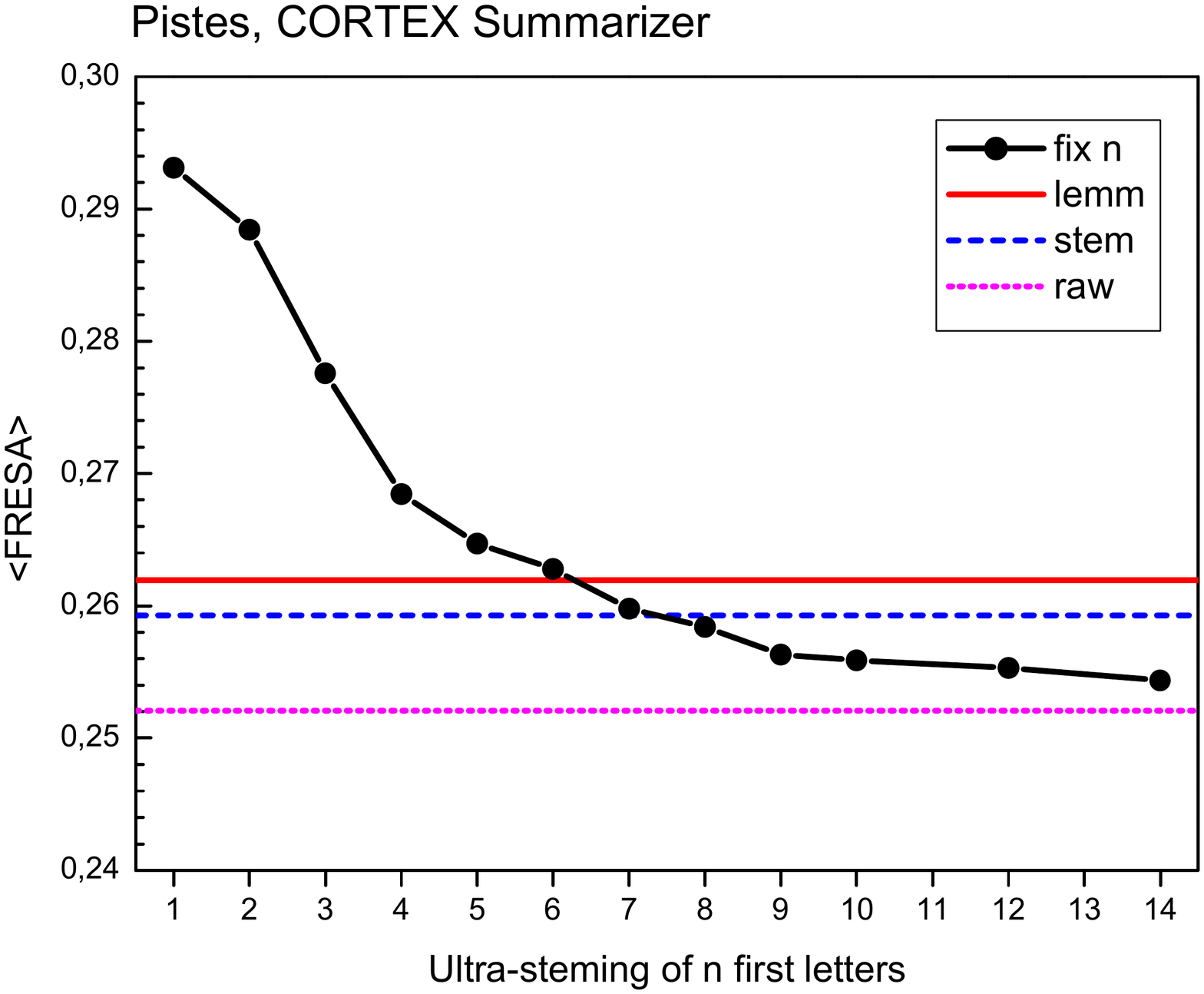}
	\caption{Scatter plot of $\langle${\sc Fresa}$\rangle$ mean vs. Ultra-stemming using $n$ first letters (corpus {\sc Pistes}, {\sc Cortex} summarizer).}
	\label{fig:pistes_fixn}
\end{figure}

Results in figure \ref{fig:pistes} show that Ultra-stemming improves the score of the three automatic summarization systems used.
In particular, the summarizer {\sc Enertex} using a stemming representation obtains a score {\sc Fresa} of 0.25928, and using {\sc Fix$_1 $} representation a score of 0.29311, 
i.e., an increase of more than 13\%.

Finally, Figure \ref{fig:pistes_fixn} shows the detailed mean score $\langle${\sc Fresa}$\rangle$ on French corpus {\sc Pistes}, 
as function of $n=1,2,...,14$ letters, using the automatic summarizer {\sc Cortex}.
As well, it shows the values {\sc Fresa} for lemmatization ({\sc Lemm}), stemming ({\sc Stem}) and plain text ({\sc Raw}).

Overall for the three languages, beyond a certain number of letters (5 for English, 7 for the Spanish and 6 for French) Ultra-stemming loses its effectiveness and lemmatization score is higher.
A view to the table \ref{tab:stats} shows that this limit has a relationship with the mean, rather than the mode of letters per word in each language.
Apparently, using Ultra-stemming is interesting when using a number of characters less than the mode of the language in question.

\section{Discussion and conclusion}
\label{sec:conclusion}

In this paper we have introduced and tested a simple pre-processing method suitable for automatic summarization text.
Ultra-stemming is fast and simple.
It reduces the size of the matrix representation, but it retains the information and charateristics of the document.
An important aspect of our approach is that it does not requires linguistic knowledge or resources which makes it a simple and efficient pre-processing method 
to tackle the issue of Automatic Text Summarization.

And what about times ?

In general, the processing times  of Ultra-stemming {\sc Fix$_1$} are shorter compared to all others methods.
Of course, processing time  depends of summarizer algorithm and pre-processing algorithm.
In general, processing time $\tau$ is function of:
$$
 \tau = \textrm{time(filtering)+time(normalization)+time(summarizer)}
$$

In our experiments, time(filtering) is independent of the summarizers and generally, filtering algorithm is very fast.
The time(normalization) depends on algorithm used (stemming, lemmatizaton) and/or extern resource (dictionary of lemmatization).
The time(summarizer) is intrinsic to each summarizer system.

By example, {\sc Cortex} is a very fast summarizer with $O(\log \rho^2)$ (where $\rho = P \times N$), and processing times for {\sc Stemming}, {\sc Raw} and {\sc Fix$_1$} are close.
In other hand, {\sc Enertex} summarizer has a complexity of $O(\rho^2)$, then it needs more time to process the same corpus. 
In this case, Ultra-stemming is a very interesting alternative to summarize long corpora.
Table \ref{tab:times} shows processing times for each corpus, 
following the normalization method for {\sc Cortex}, {\sc Artex} and {\sc Enertex} summarizers.
All times are measured in a 7.8 GB of RAM computer, Core i7-2640M CPU @ 2.80GHz $\times$ 4 processor, running under 32 bits GNU/Linux (Ubuntu Version 12.04). 

\begin{table}[H]
\centering
\begin{tabular}{l|ccc|c}
	\hline
 \bf	Summarizer	& & \bf Corpus & &  \bf Time \\ 
\hline\hline
		\sc {\bf Cortex} & \textbf{DUC'04} & \textbf{Medicina Cl\'inica} & \textbf{Pistes} & \bf Mean (All)\\
	\hline
		\textsc{Lemmatization} & 0.80' 	& 2.88'  	& 1.13' 	& 1.60'\\
		\textsc{Stemming}      & 0.40' 	& 0.26'  	& 0.53' 	& 0.54'\\
		\textsc{Raw}           & 0.33' 	& 0.26'  	& 0.41' 	& 0.40'\\
		\textsc{fix$_1$}       & 0.31' 	& 0.26'  	& 0.38' 	&\bf 0.32'\\	
	\hline\hline
		\bf Artex		  & \textbf{DUC'04} & \textbf{Medicina Cl\'inica} & \textbf{Pistes} & \bf Mean (All)\\
	\hline
		\textsc{Lemmatization} & 1.71' 	& 3.10'  	& 2.70' 	& 2.50'\\
		\textsc{Stemming}      & 1.35' 	& 0.40'  	& 2.11' 	& 1.29'\\
		\textsc{Raw}           & 1.30' 	& 0.38'  	& 2.13' 	& 1.27'\\
		\textsc{fix$_1$}       & 0.41'	& 0.28'  	& 0.51' 	& \bf 0.40'\\	
	\hline\hline
		\bf Enertex		  & \textbf{DUC'04} & \textbf{Medicina Cl\'inica} & \textbf{Pistes} & \bf Mean (All)\\
	\hline
		\textsc{Lemmatization} & 9.25' 	& 3.38'  	& 18.63' 	& 10.42'\\
		\textsc{Stemming}      & 9.28' 	& 0.75'  	& 18.38' 	& 9.47'\\
		\textsc{Raw}           & 9.16' 	& 0.73'  	& 20.76' 	& 10.22'\\
		\textsc{fix$_1$}       & 3.93'	& 0.46'  	& 8.35' 	& \bf 4.25'\\	

	\hline
\end{tabular}
\caption{Statistics of processing times (in minutes) of three summarizers over three corpora.}
\label{tab:times}
\end{table}


Clearly, the lemmatization of a large dictionary is the most time-consuming strategy. 
This is notable in the Spanish corpus, using a 1.3M dictionary entries. 
Lemmatization is at the same time, the strategy that produces the best results after the Ultra-stemming (Fix$_n$ with $n=1...4$ letters). 
In the case of Artex summarizer, the gain in time is dramatic, going from 2.50' using lemmatization to 0.40 using Fix$_1$, i.e. a gain of 625\%. 
This gain is 500\% for Cortex and 245\% for Enertex.

From our point of view, the Ultra-stemming of $n$ letters has three important advantages:
\begin{enumerate}
	\item A reduction of the space and the calculation time of automatic summarization algorithms based on the vector space model.
	\item Improving of summary content, when using $n <$ mode in letters per word of each language.
	\item Applications on resource sparse languages.
Typically $\pi$ languages where no lemmatizers, stemmers or parsers, neither corpora nor native linguist available, the Ultra-stemming can be an attractive alternative for automatic document summarizers.
\end{enumerate} 

Summarization using the Ultra-stemming representation for sentence scoring, improve the identification of most relevant sentences from documents.
The results obtained on corpora in English, Spanish and French prove that Ultra-stemming can achieve good results for content quality.
%
Tests with other corpora (DUC evaluation campaigns, TAC, INEX, etc.) in mono-and multi-document guided by a subject, and $\pi$ languages 
(Nahuatl, Maya, Somali, Interlingua, etc.) using content evaluation with or without reference summaries still in progress.


\label{sect:bib}
\bibliographystyle{plain}
\bibliography{biblio}

\end{document}